\documentclass[Physsubmission, Phys]{SciPost}

\hypersetup{
    colorlinks,
    linkcolor={red!50!black},
    citecolor={blue!50!black},
    urlcolor={blue!80!black}
}

\usepackage[bitstream-charter]{mathdesign}
\DeclareSymbolFont{usualmathcal}{OMS}{cmsy}{m}{n}
\DeclareSymbolFontAlphabet{\mathcal}{usualmathcal}

\usepackage{xcolor}

\usepackage{lineno}
\usepackage{amsmath}
\usepackage{sidecap}

\begin{document}

\begin{center}{\Large \textbf{
Nuclear modification factors of strange mesons measured by PHENIX\\
}}\end{center}

\begin{center}
A. Berdnikov\textsuperscript{1},
Ya. Berdnikov\textsuperscript{1},
D. Kotov\textsuperscript{1},
D. Larionova\textsuperscript{1},
Iu. Mitrankov\textsuperscript{1},
M. Mitrankova\textsuperscript{1} and
V. Borisov\textsuperscript{1*}
\end{center}

\begin{center}
{\bf 1} Peter the Great Saint-Petersburg Polytechnic University, St.Petersburg, Russia
\\

* borisov\_vs@spbstu.ru
\end{center}

\begin{center}
\today
\end{center}

\definecolor{palegray}{gray}{0.95}
\begin{center}
\colorbox{palegray}{
  \begin{tabular}{rr}
  \begin{minipage}{0.1\textwidth}
    \includegraphics[width=30mm]{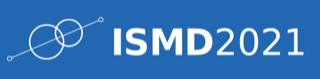}
  \end{minipage}
  &
  \begin{minipage}{0.75\textwidth}
    \begin{center}
    {\it 50th International Symposium on Multiparticle Dynamics}\\ {\it (ISMD2021)}\\
    {\it 12-16 July 2021} \\
    \doi{10.21468/SciPostPhysProc.?}\\
    \end{center}
  \end{minipage}
\end{tabular}
}
\end{center}

\section*{Abstract}
{\bf


The question of the existence and properties of the quark-gluon plasma (QGP) nowadays are the subject of detailed systematic study.
Particles that contain strange quarks can be considered as a great tool to study flavor dependence of the parton energy loss in the QGP and strangeness enhancement.
In this paper, we present the most recent PHENIX results on nuclear modification factors of $K^{\pm}$, $K^{*0}$, $\phi$ mesons as a function of $p_T$ and the number of participants in $p$+$p$, $p$+Al, $^3$He+Au, Cu+Cu, Cu+Au, Au+Au, and U+U collisions at top RHIC energies. 
The light hadron production at high-$p_T$ is apparently independent on quark content.
The coalescence mechanism might be an answer for strangeness and baryon enhancement at moderate $p_T$. 
}

\vspace{10pt}
\noindent\rule{\textwidth}{1pt}
\tableofcontents\thispagestyle{fancy}
\noindent\rule{\textwidth}{1pt}
\vspace{10pt}

\section{Introduction}

For the decades one of the most important goals of the high energy physics was investigation of matter which filled the early Universe, so-called quark-gluon plasma (QGP)~\cite{QGP}. 
In laboratory conditions the QGP can be studied with relativistic ion collisions.
The strangeness production was originally proposed as a signature of QGP formation ~\cite{StrangenessEnhancenment} and still remains experimentally popular observable since strange hadrons are produced abundantly in the QGP and can be measured over a large kinematic domain. 
The comparison of the (hidden)strange particles production to the production of hadrons that contain only first-generation quarks is a good tool to study such QGP effects as strangeness enhancement~\cite{StrangenessEnhancenment}, recombination~\cite{Recombination} and radial flow~\cite{RadialFlow} in intermediate transverse momentum ($p_T$) range and the flavor dependence of parton energy loss~\cite{JetQuenching} in high $p_T$ range.

Nowadays it is generally accepted that the QGP is formed in the relativistic heavy ion collisions such as Cu+Cu, Cu+Au, Au+Au at $\sqrt{s_{_{NN}}}$ = 200 GeV, and U+U at $\sqrt{s_{_{NN}}}$ = 193 GeV~\cite{PhidAuCuCuAuAu, PhiCuAuUU}. 
Measurements of strangeness production in large collision systems provide an opportunity to enrich and expand the understanding of the QGP properties. 

However, such QGP characteristic as minimal conditions sufficient for its formation is still under investigation in a broad set of geometry controlled small collision systems.
Recent PHENIX results indicate that the energy density in $p$/$d$/$^3$He + Au collisions at $\sqrt{s_{_{NN}}}$ = 200 GeV is ample for the creation of the deconfined state of matter in contrast to $p$+Al collisions at the same energy~\cite{QGP_min_conditions}. 
Minimal condition to form the QGP might be in between $p$ + Al and $p$ + Au collisions at $\sqrt{s_{_{NN}}}$ = 200 GeV.
However, the signatures of the QGP were observed in $p$ + Al collisions at $\sqrt{s_{_{NN}}}$ = 200 GeV at forward and backward rapidities~\cite{pAl_f_b}. 
Thus, the experimental determination of the QCD phase transition critical point require further scrutiny.

This paper presents the study of strangeness production from small to large collision systems measured by PHENIX \cite {PHENIXoverview} at top RHIC \cite{RHIC} energies. The nuclear modification factors of $K^{\pm}$, $K^{*0}$, $\phi$ mesons were measured as a function of $p_T$ and the number of binary nucleon-nucleon collisions in $p$+$p$, $p$+Al, $^3$He+Au, Cu+Cu, Cu+Au, Au+Au, and U+U collisions. 
The obtained quantities were compared to previous non-strange hadron production results to reveal the QGP effects.

\section{$R_{AB}$ results in small systems collision}
\begin{figure}[h]
\centering
\includegraphics[width=0.72\textwidth]{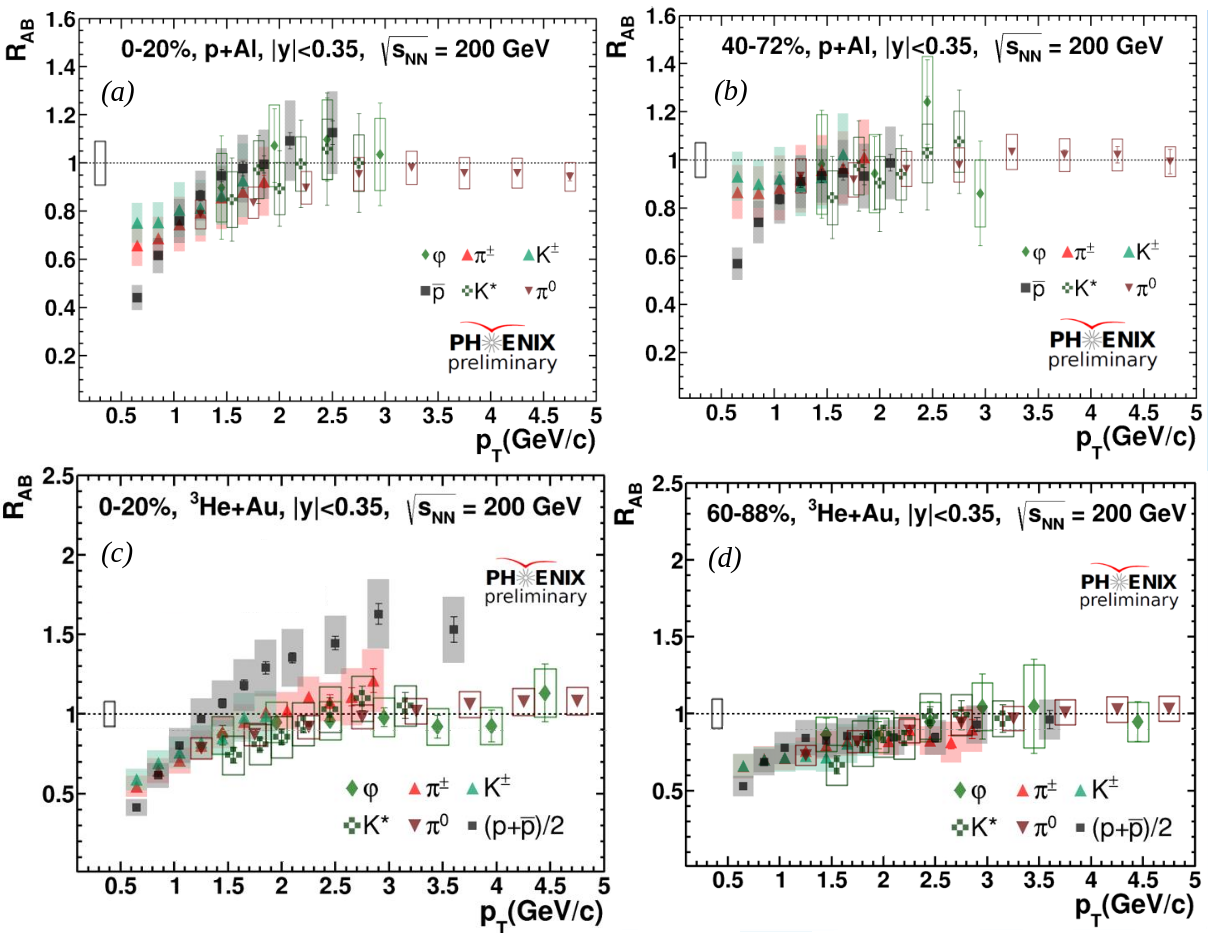}
\caption{The nuclear modification factors of $\pi^\pm$, $\pi^0$, $K^{\pm}$, $K^{*0}$, $p$($\Bar{p}$), and $\phi$ for the most central (0-20\%) ({\it a, c}) and the most peripheral (40-72\%) ({\it b}) (60-88\%) ({\it d}) $p$+Al and $^3$He+Au collisions at $\sqrt{s_{_{NN}}}$ = 200~GeV. Here and below error bars and rectangles correspond to statistical and systematic uncertainties respectively.}
\label{fig:small}
\end{figure}

Medium-induced effects on light hadron production are studied using the nuclear modification factor \cite{RAB}:

\[	 R_{AB}= \frac{d^2N_{AB}/dydp_T}{N_{coll} \cdot d^2N_{pp}/dydp_T}\]

 where $d^2N_{AB}/dydp_T$ ($d^2N_{pp}/dydp_T$) is the per-event yield of particle production in $A+B$ ($p+p$) collisions and $N_{coll}$ is a number of binary nucleon-nucleon collisions.
 
The comparison of the nuclear modification factors measured for different light hadrons ($\pi^{\pm}$, $\pi^0$, $K^{\pm}$, $K^{*0}$, $p$($\Bar{p}$), and $\phi$) are shown in the Fig. \ref{fig:small} for the most central  (0-20\% centrality class \cite{CentralitySelection}) and the most peripheral $p$ + Al (40-72\%) and $^3$He+Au (60-88\%) collisions at $\sqrt{s_{_{NN}}}$ = 200~GeV. 
In central $^3$He+Au collisions protons yields are enhanced relatively to the binary scaled yields in $p$ + $p$ collisions, while all mesons $R_{AB}$ independently of quark content lie on the same curve. 
However, all light hadron production show conformity in $p$ + Al collisions. 
This result might draw an assumption that coalescence mechanism~\cite{Recombination} plays a crucial role for light hadron production  $^3$He+Au collisions at $\sqrt{s_{_{NN}}}$ = 200 GeV at midrapidity ($\eta$ < 0.35), whereas its contribution to hadronization in $p$ + Al collisions is insignificant. 
In the most peripheral collisions in both systems there is no significant modification of all light hadron production.

\section{$R_{AB}$ results in heavy ion collisions}

\begin{figure}[h]
\centering
\includegraphics[width=0.9\textwidth]{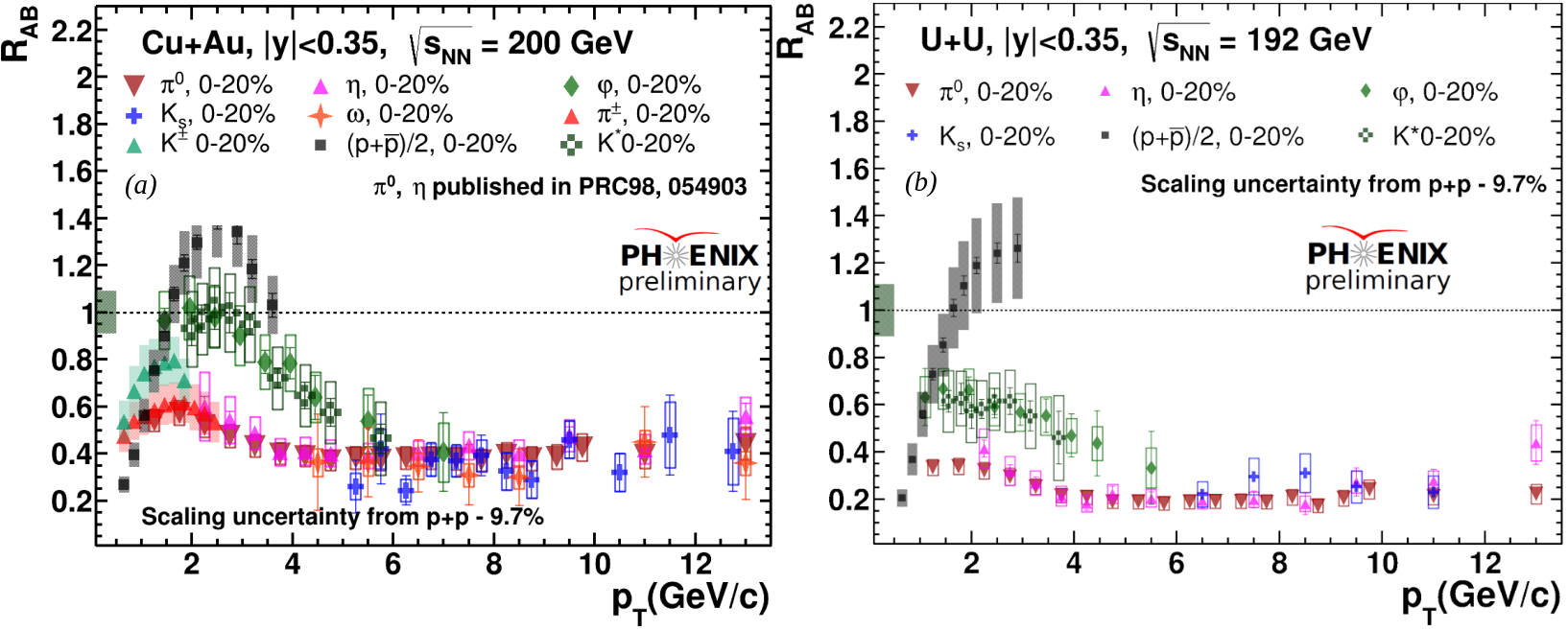}
\caption{The nuclear modification factors of light hadrons for the most central (0-20\%) Cu + Au collisions at $\sqrt{s_{_{NN}}}$ = 200~GeV ({\it a}) and U + U collisions at $\sqrt{s_{_{NN}}}$ = 193~GeV ({\it b}). }
\label{fig:heavy}
\end{figure}

In the Fig.~\ref{fig:heavy}~$R_{AB}$ of $K^{*0}$, $\phi$ mesons and charged hadrons in comparison to recently published $\pi^0$ and $\eta$ results~\cite{PiEtaCuAu, PiEtaUU} in the most central (0-20\%) Cu+Au and U+U collisions are presented. 
The data shows an ordering at intermediate $p_T$: protons $R_{AB}$ are larger than $\phi$ and $K^{*0}$ mesons (containing strange quarks) $R_{AB}$ which are larger than light flavored mesons $R_{AB}$. 
These features of hadron production can be explained with an interplay of strangeness enhancement and recombination model~\cite{Recombination}.
At high $p_T$ a flavor independent suppression of hadron production is observed. 
This result may be the consequence of the parton energy loss in the QGP.

\begin{SCfigure}
  \caption{The integrated nuclear modification factor as a function of $N_{coll}$ for $\phi$ mesons in Cu+Cu, Cu+Au and Au+Au collisions at the energy of $\sqrt{s_{_{NN}}}$ = 200~GeV and U+U collisions at the energy of $\sqrt{s_{_{NN}}}$ = 193~GeV. The tilted error bars represent the anti-correlated uncertainty on the y and x-axis due to the $N_{coll}$ calculations.}
  \includegraphics[width=0.5\textwidth]{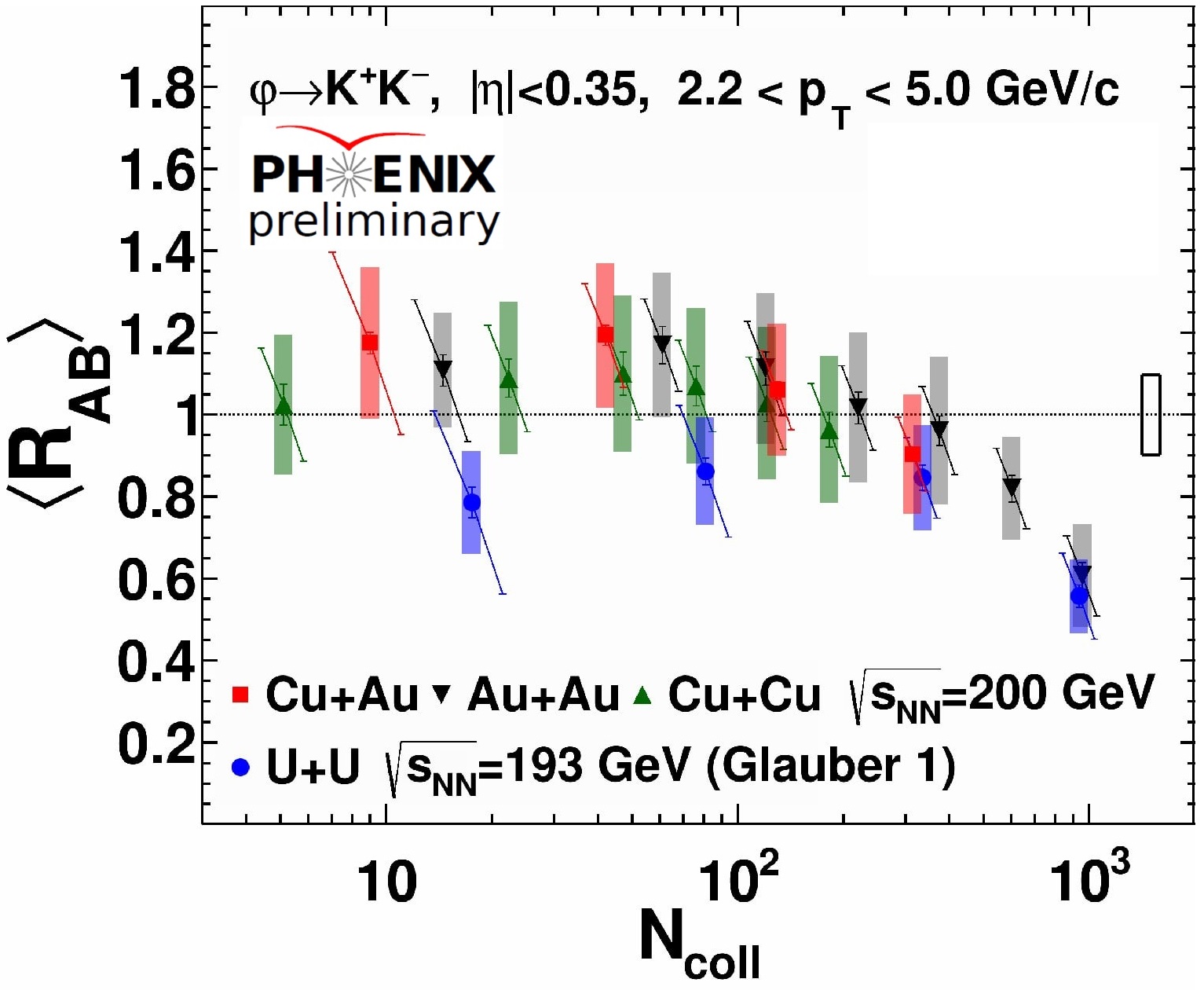}
  \label{fig:integrated}
\end{SCfigure}

In the Fig. \ref{fig:integrated} integrated $\langle R_{AB} \rangle$ of $\phi$ meson in intermediate $p_T$ range are presented. 
The strange mesons production in heavy ion collisions scales with a number of binary nucleon-nucleon collisions and seems to depend on nuclear overlap size (represented by $N_{coll}$), but not on the initial geometry of the collision system.

\section{Conclusions}
The nuclear modification factors of $K^{\pm}$, $K^{*0}$, and $\phi$ mesons were measured in p+Al, He+Au, Cu+Cu, Cu+Au, Au+Au collisions at $\sqrt{s_{_{NN}}}$ = 200~GeV, and U+U collisions $\sqrt{s_{_{NN}}}$ = 193~GeV at midrapidity as a function of $p_T$ and $N_{coll}$. 
The results were compared to non-strange mesons and (anti)protons $R_{AB}$. 

In central $^3$He + Au collisions proton production is enhanced compared to binary scaled yields in $p$ + $p$ collisions. 
Whereas no strangeness enhancement is observed within uncertainties.
The most central $p$ + Al collisions exhibit minimal or no apparent QGP effects.
Obtained results contribute to the concept that coalescence plays the predominant role in $^3$He + Au collisions, but the energy density is insufficient for observation of strangeness enhancement. 
In $p$ + Al collisions there seems to be no evidence of recombination at midrapidity.

In heavy ion (Cu + Au and U + U) collisions no flavor dependence of light hadron production at high-$p_T$ is observed. 
An obtained $R_{AB}$ ordering ($p$($\bar{p}$) $R_{AB}$ are larger than $\phi$ and $K^{*0}$ mesons $R_{AB}$ which are larger then light flavored mesons $R_{AB}$) at moderate $p_T$ in central collisions suggests that coalescence might be an answer for strangeness and baryon enhancement. 
The production of the strange mesons in heavy ion collisions scales with number of collisions and seems to depend on nuclear overlap size, but not on its geometry.


\section*{Acknowledgments}
We acknowledge support from Russian Ministry of Education and Science, state assignment for fundamental research (code FSEG-2020-0024) in the $\phi$ meson part of the analysis.



\nolinenumbers
\end{document}